\documentclass{mnras}
\usepackage{epsfig}
\usepackage{lscape}
\usepackage{rotating}
\usepackage{longtable}


\def\ls{{_<\atop^{\sim}}}

\begin{document}

\title{Diffuse Low-Ionization Gas in the Galactic Halo Casts Doubts on $z\simeq 0.03$ WHIM Detections}

\author[F. Nicastro et al.]{F. Nicastro,$^{1,2,3}$ F. Senatore,$^{1}$ A. Gupta,$^{4,5}$ S. Mathur,$^{4,6}$ 
\newauthor 
Y. Krongold,$^7$ M. Elvis,$^2$ L. Piro,$^8$ \\
$^1$Osservatorio Astronomico di Roma - INAF, Via di Frascati 33, 00040, Monte Porzio Catone, RM, Italy \\
$^2$Harvard-Smithsonian Center for Astrophysics, 60 Garden St., MS-04, Cambridge, MA 02138, USA \\
$^3$University of Crete, Heraklion, Greece \\
$^4$Ohio State University, Columbus, OH, USA \\
$^5$Columbus State Community college, Columbus, OH, USA \\
$^6$Center for Cosmology and Astro-Particle Physics (CCAPP), The Ohio State University, OH, USA \\
$^7$Instituto de Astronomia, Universidad Nacional Autonoma de Mexico, Mexico City, Mexico \\
$^8$Istituto di Astrofisica e Planetologia Spaziali - INAF, Roma, Italy}

\maketitle
\begin{abstract}
In this Letter we demonstrate that the two claims of $z\simeq 0.03$ OVII K$\alpha$ absorption lines from Warm Hot Intergalactic Medium (WHIM) along the lines of sight to 
the blazars H~2356-309 (Buote et al., 2009; Fang et al., 2010) and Mkn~501 (Ren, Fang \& Buote, 2014) are likely misidentifications of the $z=0$ OII K$\beta$ line produced 
by a diffuse Low-Ionization Metal Medium in the Galaxy's Interstellar and Circum-Galactic mediums. We perform detailed modeling of all the available high signal-to-noise 
Chandra LETG and XMM-Newton RGS spectra of H 2356-309 and Mkn 501 and demonstrate that the $z\simeq 0.03$ WHIM absorption along these two sightlines is statistically 
not required. 
Our results, however, do not rule out a small contribution from the $z\simeq 0.03$ OVII K$\alpha$ absorber along the line of sight to H~2356-309. In our model the temperature 
of the putative $z = 0.031$ WHIM filament is T$= 3\times 10^5$ K and the OVII column density is N$_{OV II} \ls 4\times 10^{15}$ cm$^{-2}$, twenty times smaller than the OVII 
column density previously reported, and now more consistent with the expectations from cosmological hydrodynamical simulations.
\end{abstract}
\begin{keywords}
Absorption Lines, ISM, Galaxy, WHIM
\end{keywords}

\section{Introduction}
According to all hydro-dynamical simulation for structure formation run in the framework of a $\Lambda-$CDM Cosmology, 
large concentrations of galaxies are the best tracers of the filamentary web of dark-matter that our local Universe is made of. 
Embedded into these filaments of shining ordinary matter (stars in galaxies), should be hidden the, still to be found, largest reservoir 
of baryons in the local Universe: the so called Warm-Hot Intergalactic Medium (WHIM: e.g Cen \& Ostriker, 2006). 
This, metal enriched (through galaxy-IGM feedback), otherwise primordial, medium should have temperatures in the range 
logT$\simeq 5-7$ (in K), and very low baryon densities $n_b \simeq 10^{-6} - 10^{-5}$ cm$^{-3}$. At such temperatures, H and He 
are mostly fully ionized (and so very difficult to detect), and the most abundant metal, oxygen, is mainly present in its stable He-like 
form: OVII. The strongest bound-bound transition of the OVII ion is the K$\alpha$ at $\lambda 21.602$ \AA, and should thus 
imprint absorption lines in the soft X-ray spectra of background objects whose lines of sight cross one or more WHIM filaments 
between us and the target. However, these lines are expected to be extremely weak. Expected OVII column densities along a 
random line of sight crossing  a WHIM filament, are N$_{OVII} \ls ${\em few}$\times 10^{15}$ cm$^{-2}$ (e.g. Cen \& Fang, 2006), 
giving rise to redshifted OVII K$\alpha$ absorption lines with EW$\ls 10$ m\AA. 
For these reasons detecting the WHIM has proven to be very challenging. The few detections so far, for the majority 
of the WHIM at logT$\simeq 6$, are either highly controversial (e.g. Nicastro et al., 2005a,b), or single-line and low 
statistical significance (e.g. Mathur, Weinberg \& Chen, 2003, Nicastro, 2010, Zappacosta, 2012). 

Perhaps the only exception that seemed to have gathered the largest consensus (e.g. Tananbaum et al., 2014) despite the 
large - compared to typical WHIM expectations - EW and associated OVII column density reported by the authors (EW=$25.8 \pm 
10.5$ m\AA), is the proposed detection of a single-line (OVII K$\alpha$) WHIM filament at the redshift of the Sculptor Wall, 
reported by Buote et al. (2009: hereinafter B09) and Fang et al. (2010: hereinafter F10) along the line of sight to the blazar H~2356-309. 
Recently, the same authors reported yet another evidence for a new OVII K$\alpha$ WHIM absorber, again at $z\simeq 0.031$, 
along the line of sight to the blazar Mkn~501 (Ren, Fang \& Buote, 2014: hereinafter RFB14). 
Indeed, the OVII K$\alpha$ resonant line shifts to $\lambda = 22.28$ \AA\ at $z=0.031$, a redshift consistent with both the 
$z=0.028-0.032$ and $z=0.03-0.04$ intervals at which the Sculptor Wall and the Hercules Supercluster cross the lines of 
sight to H~2356-309 and Mkn~501, respectively. 
For this reason B09, F10 and RFB14, identified the lines detected in the LETG and RGS spectra of these two blazars at $\lambda 
= 22.28 \pm 0.02$ \AA\ (the average and maximum semi-dispersion of the two measurements in the LETG and RGS spectra of 
H~2356-309, respectively: F10) and $\lambda = 22.31 \pm 0.02$ \AA\ (RFB14), as redshifted OVII K$\alpha$ imprinted by two 
WHIM filaments permeating the two large scale structures of galaxies, the Sculptor Wall and the Hercules Superstructures. 

In a companion paper (Nicastro et al., 2015: hereinafter N15), we present a systematic study of the cold and mildly ionized Interstellar (ISM) and Circum-Galactic 
(CGM) medium of our Galaxy through the modeling of the OI K$\alpha$ ($<\lambda> = 23.52 \pm 0.03$ \AA), OII K$\alpha$ ($<\lambda> = 23.35 \pm 0.03$ \AA) 
and OII K$\beta$ ($<\lambda> = 22.29 \pm 0.02$ \AA) absorption lines imprinted by our Galaxy's ISM/CGM in the spectra of two distinct samples of Galactic and 
extragalactic sources. 
In particular, the OII K$\beta$ transition, firstly identified by Gatuzz et al., (2013a,b), is the weak (oscillator strength $f_{OII K\beta} = 0.026$, Behar, private 
communication) $1s^2 2s^2 2p^3 \rightarrow 1s 2s^2 2p^4$ inner shell transition of the N-like (7 electrons) ion of oxygen. This line is hinted at in 9 of the 20 Galactic X-ray Binary (XRB) 
spectra and 8 out of the 29 AGN spectra of the N15 sample, and has an average rest-frame wavelength of $<\lambda> = 22.29 \pm 0.02$ \AA. 
The 8 AGNs of the N15 sample, in whose spectra the OII K$\beta$ absorption line is hinted at, include the two blazars H~2356-309 ($z=0.165388$, Jones et al., 2009) and Mkn~501 
($z=0.033663$, Falco et al., 2000), for which the same line had been instead identified as intervening $z\simeq 0.03$ OVII K$\alpha$ absorption tracing WHIM filaments (B09, F10 and RFB14). 

Here we re-analyze all the available high resolution X-ray spectra of the two sightlines towards H~2356-309 and Mkn~501 and demonstrate that, even for these two sightlines, 
the most likely identification of the $\lambda \simeq 22.30$ \AA\ line is indeed that of a $z=0$ OII K$\beta$ transition imprinted by a large amount of Low-Ionization Metal Medium 
(LIMM) that permeates the halo of our Galaxy, at large distances from the Galactic plane and perhaps up to the Galaxy's CGM (N15). This, at least for the line of sight 
to H~2356-309 (the only one for which both the K$\alpha$ and K$\beta$ transitions of OII are detectables) does not completely rule out a possible contribution, at exactly the same 
wavelengths, by redshifted OVII K$\alpha$, but dramatically limits it to an OVII column density more than ten times lower than that claimed by B09 and F10. 

\section{Data Reduction and Analysis}
We reduced and analyzed archival HRC-LETG and RGS data of the two blazars H~2356-309 and Mkn~501. 
The data of H~2356-309 include 11 HRC-LETG observations and 9 XMM-RGS observations, while Mkn~501 has only 6 XMM-RGS observations. 
All data were reduced with the latest versions of the XMM-{\em Newton} and {\em Chandra} data reduction and analysis softwares (``Science Analysis System'', Gabriel 
et al., 2004 - SAS - v. 13.5.0, and ``Chandra Interactive Analysis of Observation'', Fruscione \& Siemiginowska, 1999 - CIAO - v. 4.6.1) and calibrations 
(automatically set according to the given observation, for XMM-{\em Newton} data, and CALDB v. 4.6.2 for {\em Chandra}). 
We followed the appropriate XMM-{\em Newton} and {\em Chandra} data reduction/analysis threads and documentation to extract ``cleaned'' (excluding periods of 
high background) LETG and RGS spectra and responses for all observations and co-added their spectra to maximize the SNRE ({\bf we also checked, a-posteriori, that all the 
results of our fitting procedures to the co-added spectra of H~2356-309 and Mkn~501 where confirmed by simultaneous fitting of the individual spectra of the two 
sources}). 

The final spectra of our two targets total 193 ks (Mkn~501, XMM-RGS), 649 ks (H~2356-309, XMM-RGS) and 466 ks (H~2356-309, HRC-LETG). 

\subsection{Spectral Fitting}
We used the fitting package {\em Sherpa}, in CIAO (Freeman, Doe \& Siemignowska, 1999)  to perform spectral fitting of the RGS and LETG spectra of our targets, 
in the 6-30 \AA\ (RGS) and 6-50 \AA\ (LETG) intervals, with the main aim to disentangle local (i.e. $z=0$) absorption from our own Galaxy's ISM/CGM, from intervening 
absorption from putative WHIM filaments along the lines of sight to our two targets. 

Following the procedure described in N15, we first modeled the continuum of our targets with a power-law (model {\em xspowerlaw} in {\em Sherpa})  
attenuated by neutral absorption (the XSPEC-native model {\em xstbabs} in {\em Sherpa}: Wilms, Allen \& McCray, 2000). 
The best-fitting pure-continuum statistics for our three spectra are $\chi^2_r(dof) = 0.94(959)$ (LETG spectrum of H~2356-309), $\chi^2_r(dof) = 
1.05(3567)$ (RGS1+RGS2 spectra of H~2356-309) and $\chi^2_r(dof) = 1.03(3818)$ (RGS1+RGS2 spectra of Mkn~501). 

\begin{figure} 
\begin{center}
\includegraphics[width=84mm]{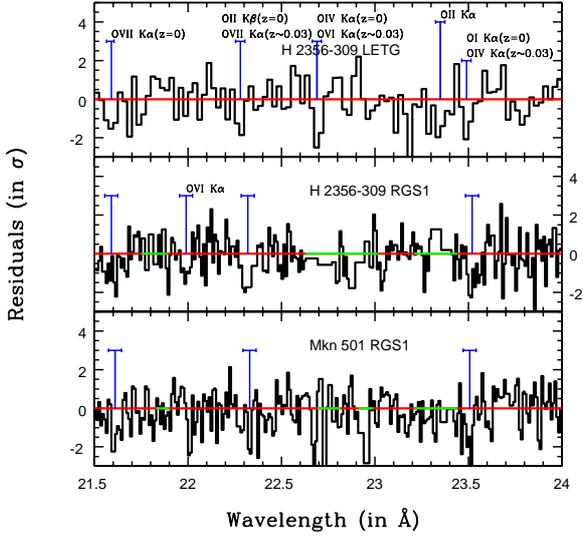}
\end{center}
\caption{ 21.5-24 \AA\ portions of the best-fitting pure-continuum model residuals (in standard deviations) of the LETG and RGS1 (the only RGS available in this spectral range) 
spectra of H~2356-309 (top and middle panel, respectively) and the RGS1 spectrum of Mkn~501 (bottom panel). The green segments in the RGS panels (middle and bottom) 
mark the positions of the RGS1 instrumental features (bad pixels). Blue vertical segments mark the best-fitting positions of the unresolved absorption lines identified in our 
analysis, while the blue horizontal segments show the FWHM of the instrument LSF.}
\label{res}
\end{figure}

\noindent
Figure \ref{res} shows the 21.5-24 \AA\ portions of the best-fitting pure-continuum model residuals (in standard deviations) of the 
LETG and RGS spectra of H~2356-309 and MKn~501. In the RGS panels, the green segments mark the spectral intervals where the presence of RGS instrumental features (bad pixels) 
hamper the search for unresolved absorption lines. 

Once a reasonable description of the broad-band continuum was obtained, we then proceded to search for, and model, the possible presence of unresolved 
(FWHM$< 950$ km s$^{-1}$ and FWHM$< 800$ km s$^{-1}$ at $\lambda=22$ \AA, for the RGS and the LETG, respectively) absorption lines in the 18-24 \AA\ band. 
To first identify candidate unresolved absorption lines and then confirm their identification in this spectral range we followed the procedure described in 
Nicastro et al.  (2013). 
This procedure yielded the results summarized in Table 1 (line IDs are also labeled in Figure \ref{res}). 
For all the lines, we report only statistical errors at the the 1$\sigma$ significance level. 

\begin{table} 
\caption{\bf Absorption Lines in the 18-24 \AA\ Spectra of H~2356-309 and Mkn~501: Model A}
\begin{tabular}{|cccc|}
\hline
$\lambda$ & Equivalent & Local ID & $z\simeq 0.03$ ID \\
& Width & & \\
in \AA\ & in m\AA\ & & \\ 
\hline
\multicolumn{4}{|c|} {H~2356-309 HRC-LETG} \\
\hline
$23.49 \pm 0.01$ & $20 \pm 8$ & OI K$\alpha$ & OIV K$\alpha$ \\
$23.35^{+0.01}_{-0.02}$ & $27 \pm 12$ & OII K$\alpha$ & N/A \\
$22.69 \pm 0.01$ & $26 \pm 9$ & OIV K$\alpha$ & OVI K$\alpha$ \\
$22.28^{+0.02}_{-0.01}$ & $18 \pm 9$ & OII K$\beta$ & OVII K$\alpha$ \\
22.02 (Frozen) & $< 30$ (at 3$\sigma$) & OVI K$\alpha$ & N/A \\
$21.59 \pm 0.02$ & $18 \pm 8$ & OVII K$\alpha$ & N/A \\
$^a$19.21 (Frozen) & $< 20$ (at 3$\sigma$) & N/A & OVII K$\beta$ \\
\hline 
\multicolumn{4}{|c|} {H~2356-309 RGS} \\
\hline
$23.52 \pm 0.01$ & $17 \pm 5$ & OI K$\alpha$ & OIV K$\alpha$ \\
$22.32 \pm 0.02$ & $17 \pm 6$ & OII K$\beta$ & OVII K$\alpha$ \\
$21.99 \pm 0.02$ & $8 \pm 5$ & OVI K$\alpha$ & N/A \\
$21.60^{+0.02}_{-0.03}$ & $14 \pm 5$ & OVII K$\alpha$ N/A \\
$^a$19.25 (Frozen) & $< 15$ (at 3$\sigma$) & N/A & OVII K$\beta$ \\
\hline 
\multicolumn{4}{|c|} {Mkn~501 RGS} \\
\hline
$23.507^{+0.008}_{-0.005}$ & $25 \pm 6$ & OI K$\alpha$ & OIV K$\alpha$ \\
$22.33 \pm 0.02$ & $14 \pm 5$ & OII K$\beta$ & OVII K$\alpha$ \\
$22.02^{+0.01}_{-0.02}$ & $11 \pm 5$ & OVI K$\alpha$ & N/A \\
$21.61 \pm 0.01$ & $16 \pm 5$ & OVII K$\alpha$ N/A \\
$^a$19.26 (Frozen) & $< 17$ (at 3$\sigma$) & N/A & OVII K$\beta$ \\
\hline
\end{tabular} \\
$^a$ For each spectrum, OVII K$\beta$ 3$\sigma$ upper limits are given at the redshift of the putative OVII K$\alpha$ WHIM line, as derived from the best-fitting 
position of the line at $\lambda = 22.28-22.33$ \AA. 
\end{table}

\subsection{Results for H~2356-309}
The 18-24 \AA\ LETG spectrum of H~2356-309 shows the presence of 5 unresolved absorption lines, which can all be identified as $z=0$ local transitions imprinted by 
a wide ionization range of oxygen ions in our Galaxy's multi-phase ISM. 
In particular, the LETG data of H~2356-309 hint at the OI K$\alpha$ (2.5$\sigma$), OII K$\alpha$ (2.3$\sigma$) and K$\beta$ 
(2$\sigma$), OIV K$\alpha$ (2.9$\sigma$) and OVII K$\alpha$ (2.3$\sigma$) transitions (Fig. \ref{res}, top panel). 
The RGS data of H~2356-309 confirm the presence of the local OI K$\alpha$, OII K$\beta$ and OVII K$\alpha$ absorption lines (detected at statistical 
significances of 3.4$\sigma$, 2.8$\sigma$ and 2.8$\sigma$, respectively) and hint at the presence of a OVI K$\alpha$ line (only 1.6$\sigma$), but cannot {\bf safely} 
detect either the OII K$\alpha$ or the OIV K$\alpha$ lines, due to the presence of instrumental features at the relevant wavelengths (Fig. \ref{res}, middle panel, green 
horizontal segments). 
\footnote{{\bf The inclusion in the analysis of these RSS regions heavily affected by the presence of instrumental features, does not change our results and  
consequent conclusions}} 
. Two of these lines, the $z=0$ OVII K$\alpha$ and the line at $\lambda = 22.28^{+0.02}_{-0.01}$ (LETG) or $22.32 \pm 0.02$ (RGS) (Table 1) were also reported by B09 and F10 (who 
use the same LETG data that we use here), who identified the second of these lines as an OVII K$\alpha$ absorber at $z\simeq 0.031-0.033$ imprinted by an intervening WHIM 
filament at a redshift consistent with that where the Sculptor Wall concentration of galaxies crosses our line of sight to H~2356-309. 
The line at $22.69 \pm 0.02$ detected in the LETG spectrum (not detectable in the RGS spectrum because of an overlapping instrumental feature) and identifiable with local 
OIV K$\alpha$ absorption, was not reported instead by F15, though the line is clearly visible in the 20-23.5 \AA\ portion of the spectrum that FN15 show in their Fig. 4. 

\subsection{Results for Mkn~501}
The 18-24 \AA\ LETG spectrum of Mkn~501 hints at the presence of the same 4 unresolved absorption lines that are detected in the RGS spectrum of H~2356-309, and that can 
be identified as $z=0$ local transitions imprinted by OI K$\alpha$ (4.2$\sigma$), OII K$\beta$ (2.8$\sigma$), OVI K$\alpha$ (2.2$\sigma$) and OVII K$\alpha$ (3.2$\sigma$) 
transitions (Fig. \ref{res}, bottom panel). 
As for the RGS spectrum of H~2356-309, the RGS spectrum of Mkn~501 cannot detect either the local OII K$\alpha$ or OIV K$\alpha$ lines, due to the presence of instrumental 
features at the relevant wavelengths (Fig. \ref{res}, bottom panel, green horizontal segments).  

Also in the case of the RGS spectrum of Mkn~501, the $z=0$ OVII K$\alpha$ line and the line at $\lambda = 22.33 \pm 0.02$ (RGS) (Table 1) were already reported by RFB14 (using 
the same RGS data that we use here), who, again, identified the second of these lines as an OVII K$\alpha$ absorber at $z\simeq 0.032-0.035$ (consistent with the source systemic 
redshift) imprinted by an intervening WHIM filament at a redshift consistent with that where the Hercules Supercluster concentration of galaxies crosses our line of sight to Mkn~501. 

\section{Discussion}

\subsection{The $z\simeq 0.03$-WHIM versus $z=0$-LIMM Conspiracy} 
From the possible identifications listed in columns 3 and 4 of Table 1, it is evident that 3 of the 6 different lines detected (or hinted at) in the LETG and RGS spectra of H~2356-309 and 
Mkn~501 have unambiguous identifications. These are, the $z=0$ OVII K$\alpha$ (LETG and RGS), OVI K$\alpha$ (RGS only in H~2356-309, where also LETG is available, but consistent 
with the 30 m\AA\ 3$\sigma$ LETG EW upper limit at that position) and OII K$\alpha$ (LETG only, because the RGS spectrum is blocked by an instrumental feature at those wavelengths), 
and can only be imprinted by a multiphase ISM/CGM, with one or more high-ionization component producing OVI and OVII absorption and a low-ionization component producing 
OII (and OI) absorption. 

\noindent
However, the 3 remaining lines have ambiguous identification. They could all be imprinted by a multi-phase mildly ionized medium in the Galaxy's ISM/CGM (with dominant OI-IV 
ions), but could also be identifiable with $z\simeq 0.03$ transitions imprinted by hotter medium, dominated by OVI-VII ions. 
This is because the OVII K$\alpha$, OVI K$\alpha$ and OIV K$\alpha$ transitions, 
at $z\simeq 0.03$ overlap with the rest frame wavelengths of the OII K$\beta$, OIV K$\alpha$ and OI K$\alpha$ transitions, respectively, expected to be imprinted by the Galaxy's 
Low-Ionization and Medium-Ionization Metal Mediums (LIMM - e.g. N15 and references therein - and MIMM). 

To discriminate between these two possibilities for these 3 absorption lines, 
we make  use of two different versions of our PHotoionized Absorber Spectral Engine (PHASE, Krongold et al., 2003) code: one photo-ionized, for the ISM/CGM of our Galaxy 
({\em galabs} model hereinafter: see N15 for details), needed to model at least the three lines with unambiguous ISM/CGM identification, and one hybridly-ionized and optimized for 
the WHIM ({\em whimabs} model, hereinafter: see, e.g., Zappacosta et al., 2010), to attempt to model at least part of the three controversial and ambiguous lines. 

The LETG and RGS data of H~2356-309 have SNRE=13 (LETG) and 26 (RGS) at 22 \AA, much larger, when combined, than the SNRE=24 of the RGS spectrum of Mkn~501. Moreover, 
the combined LETG and RGS spectrum of H~2356-309, detects 6 different lines (5 in the LETG and 3 in the RGS, 2 of which in common with the LETG), compared to the 4 lines 
detected in the lower S/N RGS spectrum of Mkn~501. H~2356-309 is therefore the best suited target to test our physically self-consistent ISM/CGM versus WHIM models.  

\subsection{The ISM/CGM Modeling}
To verify whether all the 6 different lines detected (or hint at) in the LETG and RGS spectra of H~2356-309 could be modeled by absorption due to our own Galaxy multi-phase ISM/CGM, 
we followed the procedure we used in N15. Namely, we model simultaneously the 6-50 \AA\ LETG and the 6-30 \AA\ RGS1 and RGS2 data of H~2356-309 with a continuum model 
consisting of a powerlaw (native xspec model {\em xspowerlaw} in {\em Sherpa}) attenuated by the column of neutral Galactic gas (XSPEC native model {\em xstbabs} in {\em Sherpa}, 
and add to each of the three spectra (LETG, RGS1 and RGS2) a number of {\em galabs} components till all the lines detected at single-line statistical significance $\ge 2\sigma$ could be 
adequately modeled. For each {\em galabs} component we link all parameters to a single variable value in the 3 spectra, except the redshift that is allowed to vary by $\pm 300$ km s$^{-1}$ 
from spectrum to spectrum, to allow for line misalignment within instrument resolution elements. 
This procedure required three different {\em galabs} components: (a) a low-ionization component, with a temperature of T$\simeq 4500$ K and a {\bf poorly constrained} Doppler parameter 
$b\simeq 30-100$ km s$^{-1}$ (the LIMM: dubbed Warm Ionized Metal Medium in N15), that models well the lines at $<\lambda> = 23.51$ and 23.35 \AA, identified as OI K$\alpha$ 
(best-fitting LIMM EW$^{LIMM}_{OI K\alpha} = 20$ m\AA\ versus best-fitting Gaussian $<$EW$^{Gauss}_{OI K\alpha}> = 19 \pm 5$) and OII K$\alpha$ (EW$^{LIMM}_{OII K\alpha} = 27$ m\AA\ 
versus EW$^{Gauss}_{OII K\alpha} = 27 \pm 12$ m\AA) transitions, but is able to reproduce only half of the $\lambda = 22.30$ \AA\ line, as OII K$\beta$ (EW$^{LIMM}_{OII K\beta} = 9$ m\AA\ 
versus $<$EW$^{Gauss}_{OII K\beta}> = 18 \pm 5$ m\AA); (b) a warmer mildly ionized ISM/CGM component with T$\simeq 1.2 \times 10^4$ K and b$\simeq 20$ km s$^{-1}$ (the MIMM), 
modeling less than half of the $\lambda = 22.69$ \AA\ line with the OIV K$\alpha$ transitions (EW$^{MIMM}_{OIV K\alpha} = 11$ m\AA\ versus EW$^{Gauss}_{OIV K\alpha}=26 \pm 9$ m\AA); 
(c) a hot ISM/CGM component, with T$\simeq 1.5 \times 10^5$ K and b$\simeq 20$ km s$^{-1}$ (hereinafter High-Ionization Metal Medium: HIMM), that models about 2/3 of the 
$<\lambda = 21.60$ \AA\ line, identified as OVII K$\alpha$ (EW$^{HIMM}_{OVII K\alpha} = 10$ m\AA\ versus $<$EW$^{Gauss}_{OVII K\alpha}>= 16 \pm 5$ m\AA), but fails to model the 
OVI K$\alpha$ line hinted at a single-line statistical significance of only $1.6\sigma$ in the RGS1 spectrum. 

\begin{figure} 
\begin{center}
\includegraphics[width=84mm]{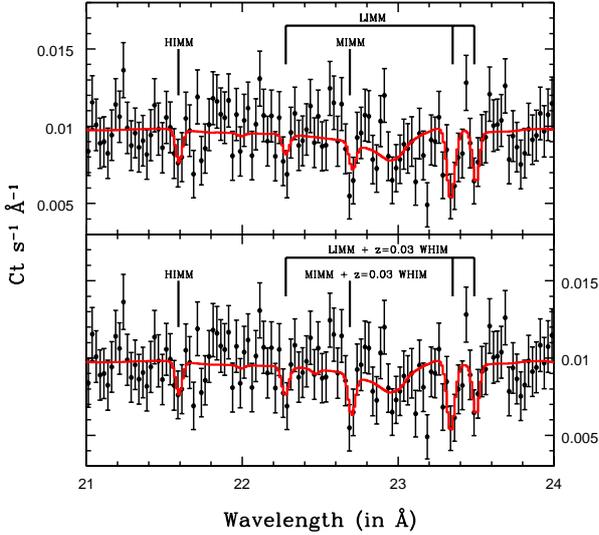}
\end{center}
\caption{21-24 \AA portions of the LETG spectrum of H~2356-309, with superimposed the best-fitting ISM/CGM {\em Model-B} including the LIMM, the MIMM and the 
HIMM (top panel)}
\label{bfmodel}
\end{figure}

Fig. \ref{bfmodel} shows the 21-24 \AA portions of the LETG spectrum of H~2356-309, with superimposed the best-fitting ISM/CGM model ({\em Model-B}) including the LIMM, 
the MIMM and the HIMM (top panel).  
The EWs of the 5 lines modeled by the three  {\em galabs} components, are consistent, within their 1$\sigma$ uncertainties, with the best-fitting EWs obtained with {\em Model-A} 
and listed in Table 1, but the best-fitting values of the OII K$\beta$, OIV K$\alpha$ and OVII K$\alpha$ are all lower than the those obtained with {\em Model-A} (single-line Gaussian fit). 
This could simply reflect the fact that the photo-ionization-equilibrium models that we use are not a sufficiently accurate description of the gas physics, but it might also indicate that 
additional components are needed. 
For example, for the unmodeled OVI K$\alpha$ line at $\lambda = 21.99 \pm 0.02$ \AA\ and the not fully modeled $z=0$ OVII K$\alpha$, a fourth, ISM/CGM phase could be present, 
with a degree of ionization in between that of the MIMM and the HIMM, while for the partly unmodeled $z=0$ OII K$\beta$ OIV K$\alpha$ an extragalactic intervening WHIM component 
may be needed, with temperature such to be dominated by OVI and OVII ions. 

We conclude that the 3-phase ISM/CGM {\em Model-B} reproduces in a statistically satisfactory way the combined LETG and RGS data of H~2356-309, but leaves some room for 
(statistically not required: $\chi^2_r(Model-B) = 1.0$ for 5332 dof) intervening $z\simeq 0.03$ WHIM absorption, filling in the 1/3 missing line opacity at the wavelengths of the 
$z=0$ OII K$\beta$ and the half missing opacity at the wavelengths of the $z=0$ OIV K$\alpha$, with the $z\simeq 0.03$ OVII K$\alpha$ and OVI K$\alpha$ transitions, respectively. 

\subsection{The ISM-WHIM Modeling}
To verify the need for an additional WHIM component, we added a {\em whimabs}' hybrid-ionization component to our best-fit ISM/CGM 3-phase {\em Model-B}, and re-fitted the data 
leaving the temperature, equivalent hydrogen column density, and Doppler parameter of the gas, free to vary and constraining its redshift to vary only between $z=0.025-0.035$ 
({\em Model-C}). 
This yielded a negligible improve in statistics (only $\Delta \chi^2 = 1.4$ for 3 additional degree of freedom over the initial 115, in the 21-24 \AA\ spectral interval), 
a best-fitting redshift of the putative WHIM filament of $<z> = 0.031 \pm 0.001$, a temperature T$_{WHIM} \simeq 3 \times 10^5$ K, a Doppler parameter $b_{WHIM} 
\simeq 100$ km s$^{-1}$ and an oxygen column density of N$_O^{WHIM} \simeq 4 \times 10^{15}$ cm$^{-2}$, in much better agreement with theoretical expectations for the WHIM 
than the implausible large oxygen column reported by F15 (N$_O \simeq 8 \times 10^{16}$ cm$^{-2}$ for their best fitting $b = 96$ km s$^{-1}$). 
The {\em whimabs} component fills in the missing opacities at the wavelengths of the $z=0$ OII K$\beta$ and OIV K$\alpha$ transitions, with $z=0.031$ 
OVII K$\alpha$ and OVI K$\alpha$ absorption, and the best-fitting {\em Model-C} molds well the 5 absorption lines detected in the LETG+RGS spectrum of H~2356-309 
(Fig. \ref{bfmodel}, bottom panel). 
Table 2 lists, for these 5 lines, the best-fitting {\em Model-C} EWs, together with the corresponding best-fitting {\em Model-A} EWs already reported in Table 1. 

\begin{table} 
\caption{\bf Comparison between {\em Model-C} and {\em Model-A} Absorption Line EWs, for the H~2356-309 Sightline.}
\begin{tabular}{|cccc|}
\hline
Local ID & $z=0.031$ ID & {\em Model-C} & {\em Model-A} \\
& & Equivalent Width & Equivalent Width \\ 
& & in m\AA\ & in m\AA\ \\ 
\hline
OI K$\alpha$ & --- & 20 & $19 \pm 5$ \\
OII K$\alpha$ & --- & 27 & $27 \pm 12$ \\
OIV K$\alpha$ & OVI K$\alpha$ & 20 & $26 \pm 9$ \\
OII K$\beta$ & OVII K$\alpha$ & 18 & $18 \pm 5$ \\
OVII K$\alpha$ & --- & 10 & $16 \pm 5$ \\
\end{tabular}
\end{table}

\subsection{Comparison with the N15 Sample} 
OII absorption is ubiquitously detected through its strongest K$\alpha$ transition, along virtually all Galactic and extragalactic lines of sight of the N15 samples. 
The 8 times weaker K$\beta$ transition is hinted at in 9 out of the 20 lines of sight and 9 out of 21 extragalactic lines of sight of the N15 samples. 
This absorption line is detected with average equivalent widths $<EW>_{OII K\beta}^{XRB} = 14 \pm 4$ m\AA\ (down to a 3$\sigma$ detectability threshold of 
$\sim 17$ m\AA) and $<EW>_{OII K\beta}^{AGN} = 18 \pm 5$ m\AA\ (down to a looser - because of the lower S/N of the AGN spectra - 3$\sigma$ detectability threshold of 80 m\AA), 
for the Galactic and extragalactic sources, respectively (N15). 

\noindent
The N15 sample of extragalactic sources includes the two blazars H~2356-309 and Mkn~501, whose LETG and RGS data we also reanalyze here. 
In our re-analysis of all the high-resolution X-ray spectra available for these two targets, for the absorption line at $\lambda\simeq 22.30$ \AA, we measure 
$EW^{Model-A} = 18 \pm 5$ and $14 \pm 5$ m\AA, for H~2356-309 and Mkn~501, respectively, fully consistent with the average EW$_{OII K\beta}$ measured 
in our samples of Galactic and/or extragalactic lines of sight in N15. 

Based on this comparison, we conclude that the $\lambda\simeq 22.30$ \AA\ absorption line observed in the spectra of H~2356-309 and Mkn~501 is likely to be produced entirely by 
OII K$\beta$ absorption. 

\section{Conclusions}
In this work we reanalyze all the available high-resolution X-ray spectra of the two blazars H~2356-309 and Mkn~501, for which claims of intervening $z\simeq 0.03$ 
WHIM absorptions were presented by B09, F10 and RFB14. 

We demonstrate that the presence of metal absorption from the diffuse LIMM in the Galaxy's ISM/CGM, casts serious doubts on these two OVII K$\alpha$ WHIM detection claims 
(B09, F10, RFB14). In particular, we show that the most likely identification for this putative $z=0.03$ OVII K$\alpha$ WHIM line, is instead that of the K$\beta$ transition of OII, 
imprinted by the Galaxy's LIMM (N15). 

For the case of H~2356-309, for which we dispose of combined {\em Chandra} LETG and XMM-{\em Newton} RGS data that allow us to detect 5 absorption lines between 
21-24 \AA, at single-line significance $>2\sigma$, we conclude that any redshifted ($z=0.031$) OVII K$\alpha$ WHIM contribution to the unavoidable presence of an OII 
K$\beta$ absorber at $<\lambda> = 22.30$ \AA, must be lower than N$_{OVII}^{WHIM} \ls 4 \times 10^{15}$ cm$^{-2}$ (for $b=100$ km s$^{-1}$), in much better agreement 
with theoretical expectations for the WHIM, than the implausible large oxygen column reported by F15 (N$_O \simeq 8 \times 10^{16}$ cm$^{-2}$ for their best fitting $b = 
96$ km s$^{-1}$).  

\section{Acknowledgements}
We thank the anonymous referee for the useful comments that helped improving the paper. 
FN and FS acknowledge support from INAF-PRIN grant 1.05.01.98.10. 



\begin{thebibliography}{}
\bibitem[\protect\citeauthoryear{Buote et al.}{2009}]{b09} 
Buote, D.A. et al., 2009, ApJ, 695, 1351: B09 
\bibitem[\protect\citeauthoryear{Cen \& Ostriker}{2006}]{cen06}
Cen, R. \& Ostriker, J.P., 2006, 650, 560 
\bibitem[\protect\citeauthoryear{Fang et al.}{2010}]{fang10}
Fang, T. et al., 2010, ApJ, 714, 1715: F10 
\bibitem[\protect\citeauthoryear{Falco et al.}{2000}]{falco00}
Falco, E. et al., 2000, ``The Updated Zwicky Catalog (UZC)'', vol 1., p. 1
\bibitem[\protect\citeauthoryear{Freeman, Doe \& Siemiginowska}{1999}]{free99}
Freeman, P., Doe S. \& Siemiginowska A., 1999, SPIE, 4477, 76 
\bibitem[\protect\citeauthoryear{Fruscione \& Siemignowska}{1999}]{fru99}
Fruscione A. \& Siemiginowska A., 1999 STIN, 9906596 
\bibitem[\protect\citeauthoryear{Gabriel et al.}{2004}]{gab04}
Gabriel C. et al., 2004, ASPC, 314, 759 
\bibitem[\protect\citeauthoryear{Gatuzz et al.}{2013a}]{gat13a}
Gatuzz, E. et al., 2013, ApJ, 768, 60 (2013a) 
\bibitem[\protect\citeauthoryear{Gatuzz et al.}{2013b}]{gat13b}
Gatuzz, E. et al., 2013, ApJ, 778, 83 (2013b) 
\bibitem[\protect\citeauthoryear{Gupta et al.}{2012}]{gup12}
Gupta, A. et al., 2012, ApJ, 756, L8 
\bibitem[\protect\citeauthoryear{Jones et al.}{2009}]{jones09}
Jones, D.H. et al., 2009, ``The 6dF Galaxy Survey Data Release 3'', vol. 1, p. 1
\bibitem[\protect\citeauthoryear{Krongold et al.}{2003}]{kro03}
Krongold, Y. et al., 2003, ApJ, 597, 832 
\bibitem[\protect\citeauthoryear{Mathur, Weinberg \& Chen}{2003}]{mathu03}
Mathur, S., Weinberg, D.H. \& Chen, X., 2003, ApJ, 582, 82
\bibitem[\protect\citeauthoryear{Nicastro et al.}{2015}]{nica15}
Nicastro, F. et al., 2015, MNRAS, submitted: N15 
\bibitem[\protect\citeauthoryear{Nicastro et al.}{2005a}]{nica05a}
Nicastro, F. et al., 2005, Nature, 433, 495 (2005a) 
\bibitem[\protect\citeauthoryear{Nicastro et al.}{2005b}]{nica05b}
Nicastro et al., 2005, ApJ, 629, 700 (2005b) 
\bibitem[\protect\citeauthoryear{Nicastro et al.}{2010}]{nica10}
Nicastro, F.  et al., 2010, ApJ, 715, 854 
\bibitem[\protect\citeauthoryear{Ren, Fang \& Buote}{2014}]{ren14}
Ren, B., Fang, T. \& Buote, D.A., 2014, ApJ, 782, L6: RFB14 
\bibitem[\protect\citeauthoryear{Tananbaum et al.}{2014}]{tana14}
Tananbaum, H. et al., 2014, RPPh, Volume 77, Issue 6, article id. 066902: rXiv:1405.7847 
\bibitem[\protect\citeauthoryear{Yao et al.}{2009}]{ya09}
Wilms, J., Allen, A., \& McCray, R., 2000, ApJ, 542, 914 
\bibitem[\protect\citeauthoryear{Zappacosta et al.}{2012}]{zap12}
Zappacosta L., Nicastro, F., Krongold, Y. \& Maiolino, R., 2012, ApJ, 753, 137 
\bibitem[\protect\citeauthoryear{Zappacosta et al.}{2010}]{zap10}
Zappacosta, L. et al., 2010, ApJ, 717, 74 
\end{thebibliography}
\end{document}